\documentclass[conference]{IEEEtran}
\IEEEoverridecommandlockouts
\usepackage{cite}
\usepackage{amsmath,amssymb,amsfonts}
\usepackage{algorithmic}
\usepackage{graphicx}
\usepackage{textcomp}
\usepackage{xcolor}
\usepackage{epstopdf}
\usepackage{multirow}
\usepackage{caption}
\usepackage{subcaption}
\usepackage[ruled,vlined]{algorithm2e}
\usepackage{mathtools}  
\usepackage{url}

\usepackage[normalem]{ulem}
\usepackage{CJKutf8}

\newcommand{\CLASSINPUTbottomtextmargin}{1in}

\begin{document}

\title{\Large \textbf{Blockchain Meets COVID-19: A Framework for Contact Information Sharing and Risk Notification System}}


\author{
    \IEEEauthorblockN{
        Jinyue Song\IEEEauthorrefmark{1},
        Tianbo Gu\IEEEauthorrefmark{1},
        Zheng Fang\IEEEauthorrefmark{1},
        Xiaotao Feng\IEEEauthorrefmark{3}
        Yunjie Ge \IEEEauthorrefmark{2},
        Hao Fu\IEEEauthorrefmark{1},
        Pengfei Hu \IEEEauthorrefmark{4},
        Prasant Mohapatra\IEEEauthorrefmark{1}
    }
    \IEEEauthorblockA{
        \IEEEauthorrefmark{1}
        Department of Computer Science, University of California, Davis, CA, USA\\
        \IEEEauthorrefmark{3}
        Department of Electrical and Computer Engineering, University of California, Davis, USA\\
        \IEEEauthorrefmark{2}
        Department of Computer Science, University of California San Francisco, USA\\
        \IEEEauthorrefmark{4}
        School of Computer Science and Technology at Shandong University\\
        Email: \{jysong, tbgu, zkfang, xtfeng, haofu, pmohapatra\}@ucdavis.edu, yge7@dons.usfca.edu, phu@sdu.edu.cn
    }
}

\maketitle


\begin{abstract}

COVID-19 is a severe global epidemic in human history. Even though there are particular medications and vaccines to curb the epidemic, tracing and isolating the infection source is the best option to slow the virus spread and reduce infection and death rates. There are three disadvantages to the existing contact tracing system: 1. User data is stored in a centralized database that could be stolen and tampered with, 2. User's confidential personal identity may be revealed to a third party or organization, 3. Existing contact tracing systems\cite{rivest2020pact}\cite{Apple_Google_virus_tracking} only focus on information sharing from one dimension, such as location-based tracing, which significantly limits the effectiveness of such systems.

We propose a global COVID-19 information sharing and risk notification system that utilizes the Blockchain, Smart Contract, and Bluetooth. 
To protect user privacy, we design a novel Blockchain-based platform that can share consistent and non-tampered contact tracing information from multiple dimensions, such as location-based for indirect contact and Bluetooth-based for direct contact. Hierarchical smart contract architecture is also designed to achieve global agreements from users about how to process and utilize user data, thereby enhancing the data usage transparency. Furthermore, we propose a mechanism to protect user identity privacy from multiple aspects. More importantly, our system can notify the users about the exposure risk via smart contracts. We implement a prototype system to conduct extensive measurements to demonstrate the feasibility and effectiveness of our system. 

\end{abstract}


\section{Introduction}\label{Intro}

Until today, coronavirus has infected more than 128 million people and caused 2.8 million deaths\cite{corvuis_worldwide}. In particular, the United States has the largest number of known infections\cite{corvuis_us}. To reduce the spread of the epidemic, people need to grasp the infection status of COVID-19 to prevent infection or take timely treatment. Every country has a different policy to share infection information, challenging to provide reliable privacy protection worldwide.

Information sharing in a centralized manner, such as MIT, Apple, and Google, announced their tracking solutions by storing users' data in cloud \cite{rivest2020pact}\cite{Apple_Google_virus_tracking}, is highly relying on users' trust. Once the personal data has been uploaded to the cloud, users cannot prevent potential data abuse.
If a comprehensive data security solution is missing, the private user data may be hacked for harmful purposes. 
As a global organization, the World Health Organization (WHO) collaborates with governments worldwide to share information and enhance epidemic prevention management. However, WHO is losing trust from some countries and cannot obtain sufficient support. Some other governments may conceal, falsely report, or hinder from reporting epidemic information, which may create a gaping security hole for global epidemic prevention\cite{Data_Security_Issue_in_Epidemic}. To this end, there is no way for individuals to share their information and protect their data privacy simultaneously.

In this work, we propose a novel framework that implements Covid-19 information sharing and exposure risk notification, which collects user data from two dimensions:
location-based indirect contact tracing and Bluetooth-based direct contact tracing. The location dimension records infections in the given locations without considering human contact. The Bluetooth dimension records the information about the direct person-to-person contact via Bluetooth technology. We utilize the Blockchain to store the data in a decentralized manner. Data consistency and user privacy can be significantly guaranteed in our system. Furthermore, smart contracts' operational functionality is implemented to avoid data abuse and opaque program execution\cite{luu2016making}, which is transparent for users about how our system collects and processes data.

More importantly, our proposed framework protects users against being identified from multiple aspects. For instance, some existing approaches give each user a unique identity to record contact among people. This paper proposes a novel method to use the Bluetooth advertisement to discover human contact without giving any unique identity to the user. Most Bluetooth-based tracing systems require the establishment of a communication channel between smartphones through Bluetooth\cite{WA_Notify}\cite{gorji2020stecc}\cite{abbas2020covid}, which may require modification of the Bluetooth communication protocol and increase the difficulty of the system developing. Also, frequent use of Bluetooth communication channels will put more pressure on smartphones' limited battery power. Our design adopts Bluetooth periodically broadcast advertisement, which does not establish communication channels, and smartphones of both users directly record mac addresses from received Bluetooth advertisements. Moreover, our design applies weak randomization of the Bluetooth addresses, which reduces the number of advertisements and the battery consumption of smartphones. We summarize the contributions as following:

\begin{itemize}
    \item We propose a comprehensive solution for utilizing Blockchain to collect and share user data to prevent the spread of Covid-19, which can ensure data security and avoid data being tampered with and stolen.
    
    \item We propose a hierarchical smart contract design to collect, process, and manage the user data, which has the executed program with global agreements for all users and dramatically enhances user data usage transparency, thereby avoiding data tamper and abuse.
    
    \item We design a weak randomized MAC method for Bluetooth to record users' direct contact and protect user identity. This method also avoids extra smartphone battery consumption and reduces data transmission congestion within the Blockchain network.
    
    \item We implement our system based on our current developed Covid-19 track system \cite{We-care} and conduct extensive experiments to demonstrate our design's feasibility and effectiveness.
\end{itemize}

\textbf{Roadmap:}
This paper is organized into the following sections: Section \ref{System_Overview} presents the system overview and system components. Section \ref{Problem_Formulation} formulizes challenging problems for our system. Section \ref{System_Design} describes the system design from the perspective of four layers and shows how these designs solve challenges. Section \ref{Evaluation} shows how we simulate and evaluate system performance. Sections \ref{Related_Work} and \ref{Conclusion} present the related work in smart contract tracing and conclusion for this paper. The last section \ref{Limitation and Future_Work} explains limitations in current work and explore research topics in the future.

\section{System Overview}\label{System_Overview}

Our system can track locations' infection status and notify users if they are exposed to these infected places. When a user updates himself infected, our system will notify other users who have been in contact with him in the last 14 days. Because of the blockchain's decentralized feature, the smart contract's notification functions are distributed on each computing node in the blockchain network, and one particular node will not become a bottleneck in the system performance. 

\subsection{Tracing User Visited Locations}

Our system provides people with a location-based tracing service, which contains two primary functions: broadcast notifications of location's infection to all visitors and allow users to check the current infection status of a specified location.

In a public environment, viruses can remain active on the surface of objects\cite{COVID_on_different_surfaces}, and float in the air in the form of aerosols\cite{ge2020possible}\cite{WHO_Protocol_for_COVID_Factors}\cite{MIT_airborne}, which means the virus can infect people by touching and breathing. Therefore, it is necessary to provide users with such services to check their destinations' infection status and whether they have been exposed to the infection environment.

In a general scenario, a user may visit many public places in his daily life, such as offices, restaurants, metro stations, or some open-air places. He can check in these locations anonymously in our proposed system by sending transactions to a designed smart contact tracing service, including his visiting date, time, and location. Then, our system will update each location's infection status based on the user's infection condition and his visiting activities. Moreover, these visiting records will be stored in decentralized blockchain databases. Also, before visiting a location, users can query the designed smart contract about the infection status of this specific location to ensure safety.

When a user reports his infected condition to our system, first, his health condition is saved locally in the smartphone, then the smart contract group embedded in our system will update the infection status of the locations or transportation that the user visited based on his visiting records. After that, another smart contract will broadcast notifications through the blockchain network to remind other users who have been in this place in the last 14 days. 

\subsection{Tracing Person to Person Contact}

Our system provides certain two functions in this service: broadcast infection notifications of Bluetooth addresses, which are provided by the smartphone applications, and allow users to check whether they have been in contact with infected patients in the past 14 days.

It is essential to trace the direct contact by Bluetooth because the detected Bluetooth advertisements have a range of 5 to 10 meters\cite{Bluetooth_range}, which indicates that there are people nearby. Moreover, close contact with patients can cause infection\cite{MIT_Sneeze}\cite{MIT_airborne}, since viruses can attach to water vapor and spread through the airborne\cite{MIT_airborne}. In this service, our design also adopts weak randomization in Bluetooth without modifying the low-level Bluetooth protocol, which reduces the communication cost and guarantees user privacy simultaneously. 

The random Bluetooth addresses are collected in daily contact between users, and these addresses are stored in users' smartphones locally for privacy protection. When a user voluntarily reports his infection, this user's smartphone application will retrieve the Bluetooth contact records in the past 14 days and package them in the transactions sent to a special smart contract. Based on the blockchain transaction gossip protocol\cite{xiao2020survey}, all users on the network can receive and forward this transaction with infection data, and then, the relevant users will check their local contact data to see if they have contact with this infected patient. Also, users can query this special smart contract about infection transactions in the past 14 days so that they can double-check their contact history for potential infection connections.

\subsection{Notifying Users for Potential Infection}

The proposed notification functionalities in our architecture are fundamental components embedded in the location-based tracing service and Bluetooth-based tracing service. Because of the integration of blockchain and smart contracts, the notification functions could become a decentralized service, which means all tasks are executed consistently by smart contracts deployed among computing nodes in the blockchain network, and no particular node will become a bottleneck in the system performance. 

Regarding the notification service, we present its four main functionalities:
\begin{itemize}
    \item Broadcast location infection to all visitors.
    \item Allow users to check the current infection status of a specified location.
    \item Broadcast anonymous infection of Bluetooth addresses from smartphone applications.
    \item Allow users to check whether they have been in contact with infected patients in the past 14 days.
\end{itemize}

\textbf{Notification about location infection: }This type of notification contains the first and second functions listed above. After users check in a location anonymously, the smart contract representing this location records users' transactions which contain their check-in information. When this place is infected, our system will broadcast and notify these users based on their check-in records that they have been to this infected place. Besides, we consider that actively querying the infection status of a location is also a type of notification service, so in our design, users can also interact with the smart contract to get the current infection status of this location as a reference for travel.

\textbf{Notification about individual infection: }This type of notification contains the remaining two functions. Based on the blockchain transaction gossip protocol\cite{xiao2020survey}, when an infected user sends transactions containing his Bluetooth addresses to our proposed specific smart contract, other users will receive and forward these transactions in the blockchain network. In this transaction propagation process, users eventually catch all transactions about infection, and they can check if their Bluetooth addresses have connections with these infection transactions.
Users can also actively query the smart contract to see if their used random Bluetooth addresses appear in the infected user's transactions in the last 14 days.

\section{System Challenges}\label{Problem_Formulation}

Regarding system performance, data security, and user privacy, we have five fundamental challenges: 1) Transaction latency needs to be minimum when the system expands. 2) Throughput needs to increase with the system's scale-up. 3) Operating costs need to remain reasonable and stable. 4) User data needs to be secured in an untrusted network environment. 5)  User's identity privacy needs to be protected when numerous Bluetooth addresses are broadcast.

\subsection{Latency Minimization} 
In our system, we define the latency as a time difference $\mathtt{\Delta time}$ about how long it takes to process a $\mathtt{Req^{checkin}}$ entirely by smart contracts and store it in the database.

 We expect to reduce latency to improve system efficiency by optimizing the design of the system, where the latency is affected by the following five factors: 1. the number of users $\mathtt{|U|}$ in the system, 2. the frequency $\mathtt{Freq}$ of requests, 3. the block size $\mathtt{|B|}$, 4. the height of the smart contract group $\mathtt{SCG.height}$, and 5. the length of the waiting queue $\mathtt{|Queue|}$ of the smart contract. The difficulty of this challenge is that when the number of smart contracts increases, our system should still efficiently process requests and transactions, which means the latency should have a stable minimization.

In the system, the total number of users and the frequency $\mathtt{Freq}$ of user's check-in determine the total number of user requests per unit time.
If the amount of $\mathtt{Req^{checkin}}$ exceeds the smart contract's processing capacity, the following $\mathtt{Req^{checkin}}$ will enter the $\mathtt{Queue}$. If the number of unprocessed requests is greater than the length of the queue, these requests will be abandoned. Then the smart contract needs to wait for other nodes to synchronize the transaction operations and data, which brings a longer latency. 

We introduce the block size $\mathtt{|B|}$ because it is one of the bottlenecks in the low-level protocol design of blockchain towards high throughput and low latency\cite{10.1007/978-3-662-53357-4_8}. The height of the smart contract group $\mathtt{SCG.height}$ is associated with our proposed hierarchical structure, where with the increase of $\mathtt{SCG.height}$, latency should maintain relatively stable and slow growth. 

We establish the latency formula:
\begin{eqnarray}
\mathtt{Latency} = \{\mathtt{|U|}, \mathtt{Freq}, \mathtt{|B|}, \mathtt{SCG.height}, \mathtt{|Queue|}, \nonumber\\\mathtt{\delta^{U,F}}, \mathtt{\delta^{|B|}}, \mathtt{\delta^{SCG.height, |Queue|}}, \mathtt{\phi}\}
\end{eqnarray}
where $\mathtt{\delta}$s are a series of transition functions to determine the latency, and $\mathtt{latency} = \mathtt{\phi(\delta^{U,Freq}, \delta^{|B|}, \delta^{SCG.height, |Queue|})}$.

\subsection{Throughput Maximization}
Throughput $\mathtt{TP}$ can intuitively show the system’s ability to process user requests. It refers to the number of user requests wholly handled in a unit of time by the system, which is affected by $\mathtt{Latency}$, packet loss rate $\mathtt{Rate^{PL}}$ and bandwidth $\mathtt{BW}$. $\mathtt{Latency}$ is affected by the five factors described in the previous section, and the $\mathtt{Rate^{PL}}$ and $BW$ depend on the network conditions at the user end. Among these three factors, only $\mathtt{Latency}$ is within the scope of system design. Similarly, when the system scales up, we need $\mathtt{TP}$ to increase accordingly. Therefore, we need to reduce $\mathtt{Latency}$, thereby increasing $\mathtt{TP}$ and improving system efficiency.

Therefore, the throughput can be defined as follows:
\begin{eqnarray}
\mathtt{TP} = \mathtt{\{Latency, Rate^{PL}, BW, \theta}\}
\end{eqnarray}
where $\mathtt{\theta}$ is the transition function with three arguments to determine the $\mathtt{TP}$.

\subsection{Minimum Operating Cost Optimization}
Operating cost is the focus of the challenges. We expect that while the system is scaled up, the operating cost is still minimized in our proposed design. The following section will present how our design provides a practical and reasonable solution for this challenge. 

The system's operating cost is the total amount of Ethereum gas\cite{Ethereum_Yellow_Paper} used in all four layers.  It measures the following four factors: location-based and Bluetooth-based contact tracing services, smart contract setup, and general operations in smart contracts. The cost of querying a database is not considered because users could interact with their local synchronized database\cite{dinh2018untangling}. Therefore, the challenge lies in whether the overhead of the two tracing services is within an acceptable range during operation and whether the maintenance of smart contracts will increase costs.

We assume that the number of users has a polynomial relationship with requests through two tracing services. The setup cost of a smart contract is fixed, but its operating costs will increase with increased users. So we have the following cost formula:
\begin{eqnarray}
\mathtt{Cost = \{Loc, Bt, Setup, \lambda\}}
\end{eqnarray}
where $\mathtt{Loc}$ represents operation cost in location-based contact tracing service, $\mathtt{Bt}$ represents cost in Bluetooth-based service, and $\mathtt{\lambda}$ is the transition function to calculate the cost combined by these three factors. 

Then, we quantify the average and variance values of the operating cost, representing the system's stability and expectation in optimal local conditions. So, we have the formulas when the system has three factors at minimal cost:
\begin{eqnarray}
\mathtt{args_{var} = \mathtt{argmin}_{var}(\lambda(Loc, Bt, Setup))}\\
\mathtt{args_{avg} = \mathtt{argmin}_{avg}(\lambda(Loc, Bt, Setup))}
\end{eqnarray}
Therefore, we can calculate the optimal arguments with a minimal system cost:
\begin{eqnarray}
\mathtt{args_{optimal} = minCost(args_{var}, args_{avg}, \zeta)}
\end{eqnarray}
where we introduce the penalty function $\zeta$ to adjust average and variance values for optimal arguments.

\subsection{Data Security Guarantee}

The challenge of storing user data in an untrusted distributed network is how to ensure the security and integrity of the data.

When an attacker tries to tamper with and broadcast a user's or a location's infection information, our system should prevent such information tampering and ensure the integrity and correctness of the data. The database synchronized with our system on each user's side should be consistent, and the results of the infection status query sent by the user should also be consistent.

Whether such data security requirements can be guaranteed depends on the choice of the database architecture in our system. In addition, we also need to ensure the correctness of the programs distributed on each computing node to avoid malicious operations. For example, if a healthy user checks in a restaurant, his transaction should not cause the restaurant's status to become infected.

\subsection{Identity Privacy Protection}

Our system uses the Bluetooth address to represent the user's identity, so the challenge is whether the weakly randomized Bluetooth address can represent the user without revealing the user's true identity.

In order to protect the user's privacy, the user's identity can neither be bound to a fixed series of numbers and symbols nor should it be deduced and matched. When a string of digital symbols binds the user's identity, this user's travel information and health status will be exposed to the dangerous public network. Even if this string of digital symbols is random, attackers can match the user's behavior pattern with other databases and deduce this user's real identity. Therefore, anonymous and fixed identity representation cannot protect user privacy.

In addition to this, we are also facing minor challenges regarding the additional overhead caused by randomizing the Bluetooth address, such as mobile phone battery consumption and increased system operating costs.

\section{System Design}\label{System_Design}

This section defines and explains our system components from the four layers' perspectives and the interactions between tracing and notification mechanisms. Also, these designs are shown to solve the three challenges mentioned in the previous section.

\subsection{System Architecture}

    \begin{figure}
    \centering
      \includegraphics[width=1\columnwidth]{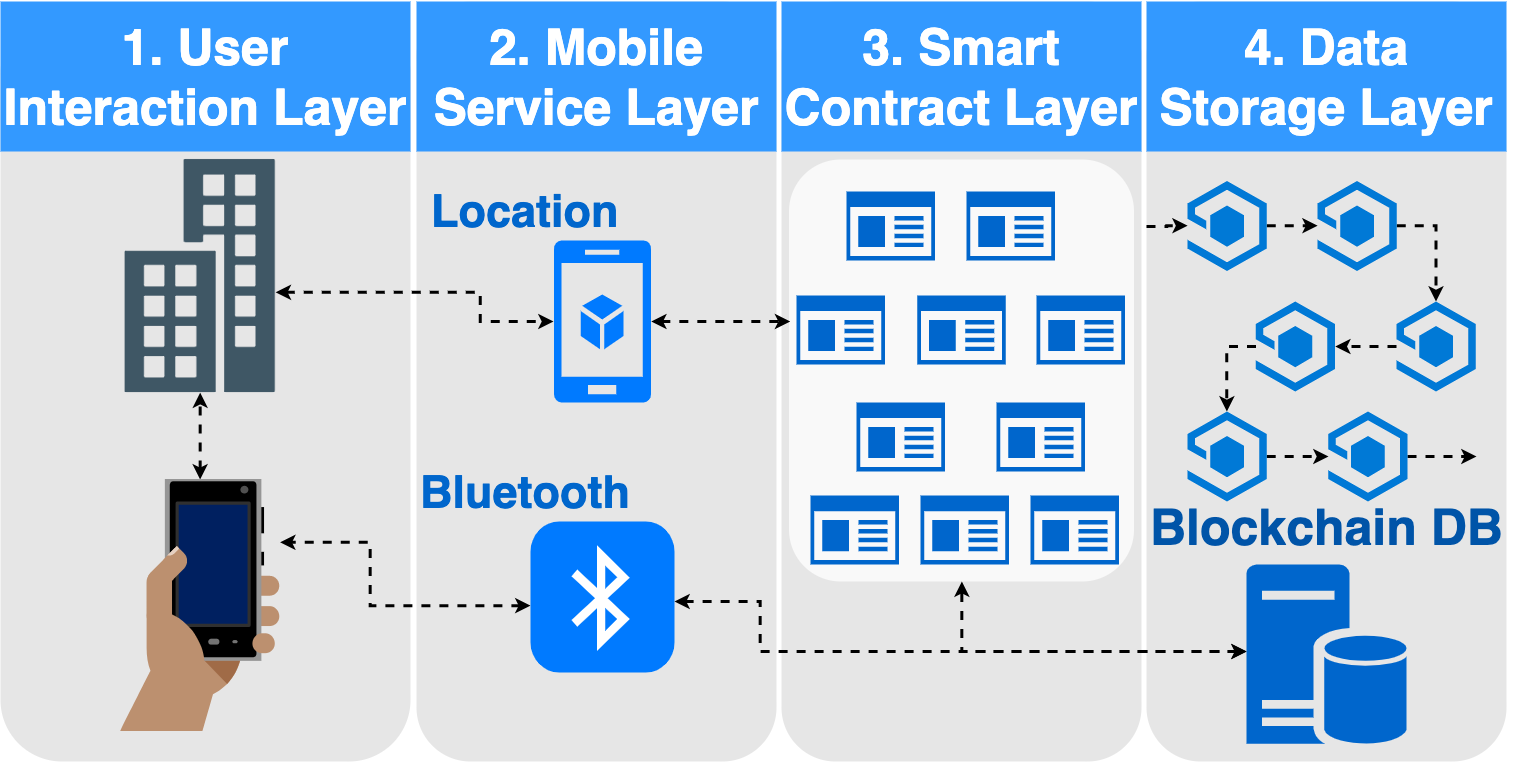}
      \caption{Blockchain enabled COVID-19 Trace and Notification Architecture.}~\label{fig:sys_arch}
      \vspace{-0.7 cm}
    \end{figure}
    
Our system contains four layers: \textbf{User Interaction Layer}, \textbf{Mobile Service Layer}, \textbf{Smart Contract Service Layer}, and \textbf{Data Storage Layer}, shown in Fig. \ref{fig:sys_arch}. 
Moreover, it provides two primary services for trace and notification: \textbf{Bluetooth-based personal contact trace service} and \textbf{location-based contact trace service}. 
The Bluetooth-based contact trace service is supported in the second layer. The location-based contact trace is coordinated by the smart contracts primarily in the third layer. Anonymous infection Bluetooth addresses and location check-in records are stored in Blockchain databases in the fourth layer. 

(a) \textbf{User Interaction Layer}\\
At the User Interaction Layer, we have two entities: user $\mathtt{U}$ and location $\mathtt{L}$. Users hold Bluetooth-enabled smartphones and have two health status types: healthy users $\mathtt{U^{normal}}$ and infected users $\mathtt{U^{infected}}$. In the second layer, users access Mobile Service to update their infection status in the system. We assume that users are always honest with their infection status.

    \begin{figure*}
    \centering
      \includegraphics[width=1.5\columnwidth]{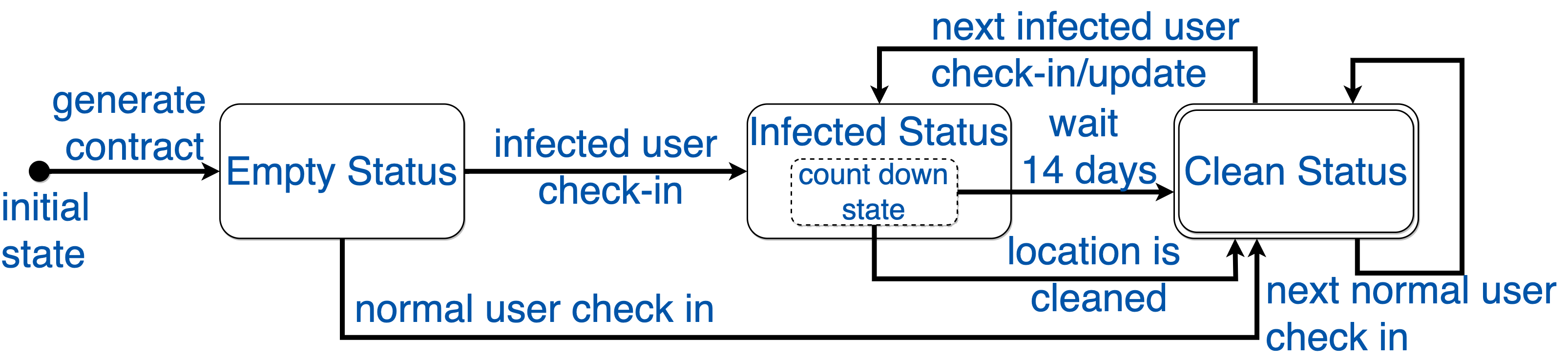}
      \caption{Location-based Contact Tracing State Machine.}~\label{fig:location_state_machine}
      \vspace{-0.5 cm}
    \end{figure*}
    
A location $\mathtt{L}$ is a public place or transportation that users often visit, such as an office, restaurant, bus, and even airplane. Location $\mathtt{L}$ also has two status types, uninfected location $\mathtt{L^{normal}}$ and infected location $\mathtt{L^{infected}}$. If an infected user $\mathtt{U^{infected}}$ visited this location, or a user updates himself to be infected after his visiting, then this location $\mathtt{L}$ would be marked as $\mathtt{L^{infected}}$ by a smart contract.

(b) \textbf{Mobile Service Layer}\\
Mobile Service Layer is the core handler in our system. It interacts directly with the other three layers and provides users with three services: \textbf{location-based tracing service}, \textbf{Bluetooth-based tracing service}, and \textbf{notification service}. 

In the \textbf{location-based tracing service}, users send check-in requests $\mathtt{Req^{checkin}}$ from their smartphone application to the smart contract that represents their visiting location. This $\mathtt{Req^{checkin}}$ contains the user's random Bluetooth MAC address $\mathtt{MacAddr}$ as a virtual identity, the visiting date $\mathtt{Date}$, and the geographical position $\mathtt{GeoPos}$. Except for these check-in requests $\mathtt{Req^{checkin}}$, the infection status of this location $\mathtt{L}$ is also affected by users who update themselves as infected. When a user updates himself as infected $\mathtt{U^{infected}}$, his smartphone application will send transactions to the corresponding smart contracts that represent his visiting locations within the past 14 days. After that, locations $\mathtt{L}$ will be marked as infected $\mathtt{L^{infected}}$  by these corresponding smart contracts.
    
The \textbf{Bluetooth-based service} is an integral part of the proposed smartphone application. It scans the surrounding Bluetooth advertisements from other users and broadcasts its advertisements with Bluetooth random $\mathtt{MacAddr}$. Then it calculates the time interval $\mathtt{TimeInterval}$ and received signal strength index $\mathtt{RSSI}$ of this interaction between users to estimate the distance between them. When a user finds himself infected, he will send transactions containing his random Bluetooth mac address in the past 14 days with infection conditions to a particular smart contract for recording. Otherwise, these contact data will save locally on the user's smartphone. 

\textbf{Notification service} provides functions for the other two services in the Mobile Service Layer. This service can remind users if they visited infected locations or if they had in contact with infected patients.

\textbf{Notification on location-based service:} 
Once receiving the infection transactions, the smart contract representing the location will update this location as infected based on the transactions' check-in data. Then it counts down 14 days from the provided $\mathtt{Date}$, and sends reply transactions as notifications to users who check in this place in the coming 14 days. Also, 14 days before this place was updated as infected, the virus may already exist, and users who had visited were in a state of possible infection. Therefore, these users will also receive notifications of the location's infections.

\textbf{Notification on Bluetooth-based service: }
Since the blockchain is built on gossip protocol\cite{vujivcic2018blockchain}\cite{xiao2020survey}, most users can be notified by receiving the infection transactions broadcast from the infected user. Their smartphone application will check local contact records and query smart contracts with the given infected $\mathtt{MacAddr}$ to alert users.

(c) \textbf{Smart Contract Service Layer}\\
This layer is the second core of our system. It contains a 4-level administrative hierarchy architecture, which is supported by smart contract groups\cite{song2020smart} to manage and maintain location infection status based on the check-in requests. Also, it processes and responses to users' infection Bluetooth transactions.

Based on this administrative hierarchy system\cite{song2020smart}, state contracts $\mathtt{Contract^{state}}$ are on the top level, followed by the county $\mathtt{Contract^{county}}$, then the city $\mathtt{Contract^{city}}$, and finally the lowest location contracts $\mathtt{Contract^{location}}$. Contracts at the lower level belong to those at the upper level. Each contract will only inherit from one superior contract instead of two different superior contracts simultaneously.

Because of smart contracts' distributed property and automatic execution property, user requests at different computing nodes will not occupy each other's computing resources, and they can synchronize the execution sequence to obtain the same execution result. In addition, the requests received by the smart contracts at the same level in the group are all sent from the higher-level smart contracts because the hierarchical structure we propose is a tree structure. Users in various regions will execute local smart contracts, because each contract of the same level is independent of each other, and a particular contract will not constrain the system's latency. So, there will be no blockage due to the increase in contracts, and this design solves the latency challenge, which means it solves the throughput challenge also. 

The $\mathtt{Contract^{location}}$ dynamically updates its infection status of the corresponding location $\mathtt{L}$ among three states: \{$\mathtt{Empty Status}$, $\mathtt{Infected Status}$, $\mathtt{Clean Status}$\} depending on the received infection transactions. If an infected user $\mathtt{U^{infected}}$ visits this location $\mathtt{L}$, or a user who has visited this location, reports that he is infected, then this location $\mathtt{L}$ is considered infected by this user $\mathtt{U^{infected}}$. When the location is cleaned, or 14 days after being infected, this location $\mathtt{L}$ is considered to be in a $\mathtt{Clean Status}$.
    
To solve the operation cost challenge and have better maintaining of the location infection status, only the coming requests $\mathtt{Req}$s trigger $\mathtt{Contract^{location}}$ to check and update the infection status, whose operation cost is linearly associated with users' request. This design ensures that users can get the latest infection status while avoiding the contract's unnecessary self-check operations, causing extra costs.

We proposed a $\mathtt{Contract^{special}}$ in Bluetooth-based tracing service to receive all $\mathtt{Tx}$s containing infected Bluetooth contact records only, which are stored in the blockchain database \cite{Ethereum_Yellow_Paper}\cite{vujivcic2018blockchain}. It supports users to synchronize these contact records based on the $\mathtt{Contract^{special}}$ address from the APIs\cite{Etherscan_API} without gas consumption. The contact history is queried locally on the user's database, which places no operating cost requirements on the system. Therefore, the third challenge, operating costs, has also been resolved.

(d) \textbf{Data Storage Layer}\\
We propose a distributed blockchain database $\mathtt{DB}$ in the Data Storage Layer. Different from centralized database\cite{le2006biomodels} and traditional distributed database\cite{lakshman2010cassandra}, this design provides a consistent and synchronized database among entities in the network, so every user and computing node can get the same complete database.

This blockchain database will store all infection related transactions $\mathtt{Tx}$ including check-in requests $\mathtt{Req}$,  users' Bluetooth contact records $\mathtt{MacAddr}$, $\mathtt{TimeInterval}$, $\mathtt{RSSI}$, and geographical position $\mathtt{GeoPos}$ of users' visit.

Current Bitcoin and Ethereum blockchain database designs have performance bottlenecks such as high computational cost and low transaction throughput\cite{blockchain}. Moreover, there are existing third-generation blockchain databases now whose performance, such as throughput, can be comparable to databases of VISA credit card companies\cite{mcconaghy2016bigchaindb}\cite{visa_throughput}. However, the third-generation blockchain database has no mature commercial development and lacks a mature platform to support the construction of smart contracts. Thus, we deploy our system on the mature platform Ethereum for simulation and evaluation.

\subsection{Location-based Contact Tracing}

This section describes the location-based contact tracking service's working principle by illustrating the entire service process, defining the entity participating in the service, and mathematically formalizing service functions.

In Fig. \ref{fig:location_state_machine}, we illustrate location-based contact tracing service with a simplified state machine.
\begin{eqnarray}
\{\mathtt{Q}, \delta, \mathtt{InitialState}, \mathtt{CleanStatus}\}
\end{eqnarray}
where we have 
\begin{itemize}
    \item $\mathtt{Q}$ = \{$\mathtt{Empty Status}$, $\mathtt{Infected Status}$, $\mathtt{Clean Status}$\}
    \item $\mathtt{\delta}$ = \{$\mathtt{Generate Contract}$, $\mathtt{Normal User Checkin}$,\\
    $\mathtt{Infected User Checkin}$, $\mathtt{Infected User Update}$,\\ 
    $\mathtt{Location is Cleaned}$, $\mathtt{Wait 14 Days}$\}
    \item $\mathtt{InitialState}$ is the Null before the contract initialization
    \item $\mathtt{CleanStatus}$ is the accepting state
\end{itemize}

First, when a user issues a check-in request $\mathtt{Req^{checkin}}$ through the location-based contact service at the second layer, if the corresponding $\mathtt{Contract^{location}}$ to this check-in location does not exist yet, the system will create one from the location's superior $\mathtt{Contract^{city}}$ and tag it as $\mathtt{Empty Status}$. If this $\mathtt{Contract^{location}}$ exists and the $\mathtt{Req^{checkin}}$ is infected, this location updates itself to the $\mathtt{Infected Status}$. Otherwise, it enters the $\mathtt{Clean Status}$. When this location is in $\mathtt{Infected Status}$, either waiting 14 days or being cleaned,it can transform to $\mathtt{Clean Status}$. Finally, in the state $\mathtt{Clean Status}$, if coming $\mathtt{Req^{checkin}}$ is infected, or a user updates to be infected, this location will be in $\mathtt{Infected Status}$. Otherwise, it remains in the current $\mathtt{Clean Status}$.

\begin{figure}
    \centering
      \includegraphics[width=1\columnwidth]{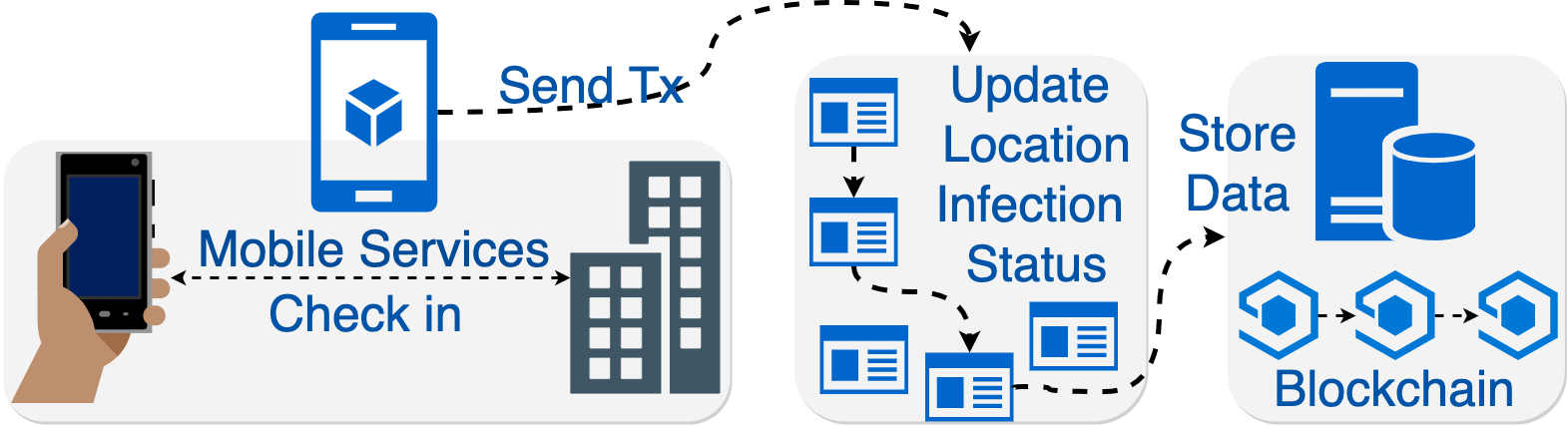}
      \caption{Location-based Contact Tracing: check-in locations, interact with smart contracts and store records in blockchain.}~\label{fig:location_contact_tracing}
      \vspace{-0.5 cm}
    \end{figure}

Fig. \ref{fig:location_contact_tracing} is an example of user $\mathtt{U}$ checking in a building. His $\mathtt{Req^{checkin}}$ includes timestamp $\mathtt{T}$, geographic position $\mathtt{GeoPos}$ and health status $\mathtt{U^{infected}}$, which is packaged in a transaction $\mathtt{Tx}$ sent to the smart contract group in the third layer. According to $\mathtt{GeoPos}$ of this building, the $\mathtt{Tx}$ is passed from the state-level $\mathtt{Contract^{state}}$ to the lowest level $\mathtt{Contract^{location}}$ representing this building. Then, due to the $\mathtt{U^{infected}}$, this $\mathtt{Contract^{location}}$ changes to $\mathtt{Infected Status}$. This $\mathtt{Tx}$ with user's check-in data is stored in the blockchain database $\mathtt{DB}$.

Also, users can acquire the infection status of a location from this service, which needs to discuss based on whether the users have the $\mathtt{Contract^{location}}$ network address locally on their smartphones.
If the user has checked in this location before or has previously queried the smart contract of this location, the corresponding $\mathtt{Contract^{location}}$ address should be saved in his smartphone application. Therefore, the user can directly send a request to the $\mathtt{Contract^{location}}$ for the infection status. If this is the first time for the user, he has to get the $\mathtt{Contract^{location}}$ network address first. With this location's $\mathtt{GeoPos}$ provided from the user, the state-level $\mathtt{Contract^{state}}$ will return this $\mathtt{Contract^{location}}$ address to him if it exists in the hierarchical group. Otherwise, $\mathtt{Contract^{state}}$ will trigger the lower level $\mathtt{Contract^{city}}$ to create a new $\mathtt{Contract^{location}}$, and return this new address to the user. Based on the location infection status, the user's smartphone application will notify him for safety. 

After receiving the location infection status from Smart Contract Service Layer, the user's smartphone can verify the response by querying his local blockchain database for the infection records. If the infection exists in this location $\mathtt{L^{infected}}$, our proposed application will alert the user.

To encourage users to access services more frequently, we proposed a check-in and query incentive mechanism. $\mathtt{Contract^{location}}$ will return more transaction fees to users when they check-in locations or query for infection status. The additional fee also supports users in utilizing mobile services such as broadcasting their transactions containing the Bluetooth contact data, check-in information, and health status.

\subsection{Bluetooth-based Contact Tracing}

In this section, we use examples and illustrations to demonstrate this Bluetooth-based contact tracing service. It involves all entities and layers, including $\mathtt{Contract^{special}}$ in Smart Contract Service Layer. 

First, the smartphone application packs users' four essential data in the transactions: time intervals that users detected each other $\mathtt{\Delta time^B}$, the detected mac addresses $\mathtt{MacAddress}$, smartphone model $\mathtt{DeviceType}$, and the received signal strength index $\mathtt{RSSI}$. Then, this application broadcasts transactions on the blockchain network, all saved in the blockchain database. For security and privacy purpose, users' Bluetooth contact records are saved locally on their smartphones. 

If a user is infected, we assume that he is honest to send $\mathtt{Tx}$s containing his contact records in the last 14 days to $\mathtt{Contract^{special}}$. Due to the blockchain gossip protocol\cite{vujivcic2018blockchain}\cite{xiao2020survey}, every participant in the network can receive and forward the infected Bluetooth $\mathtt{Tx}$s. When users' smartphone application receives transactions about infected users' status and Bluetooth addresses, it will match users' contact records with their local data and alert them if there is a matched Bluetooth address with infected patients. 

\begin{figure}
    \centering
      \includegraphics[width=1\linewidth]{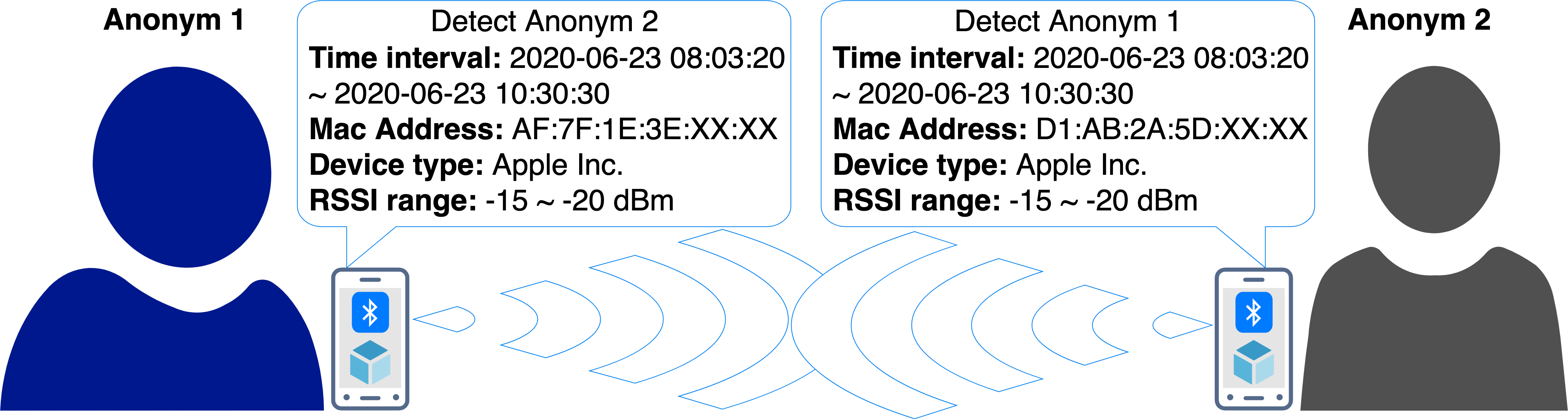}
      \caption{Bluetooth-based Person-to-Person Contact Tracing.}~\label{fig:bluetooth_contact_tracing}
      \vspace{-0.5 cm}
    \end{figure}

Fig. \ref{fig:bluetooth_contact_tracing} shows that when users are close to each other, their smartphone application scans surrounding users through Bluetooth. At the same time, it calculates the time interval $\mathtt{\Delta t^B}$ of direct contact of the users, and the range of $\mathtt{RSSI}$, and then, it records the random mac address $\mathtt{MacAddr}$, and the type of mobile device $\mathtt{DeviceType}$ of the others. After that, this application packets the four elements $\mathtt{\Delta time^B}$, $\mathtt{MacAddr}$,  $\mathtt{DeviceType}$, and $\mathtt{RSSI}$ into a transaction to broadcast in this blockchain network.

There is a scenario to address. When the smartphone application receives a transaction containing another user's infected health status and his Bluetooth mac addresses, the notification function in Bluetooth-based contact tracing will query its local database for these mac addresses and check whether the current user has a contact record with the infected user. If they have a contact history, then the application will alert the user. Then, the application will propagate this transaction again on the network to alert other users who may have contact with the infected user. 

\begin{figure*}[t]
    \centering
    \begin{subfigure}[t]{0.3\textwidth}
        \centering
        \includegraphics[height=1.2in]{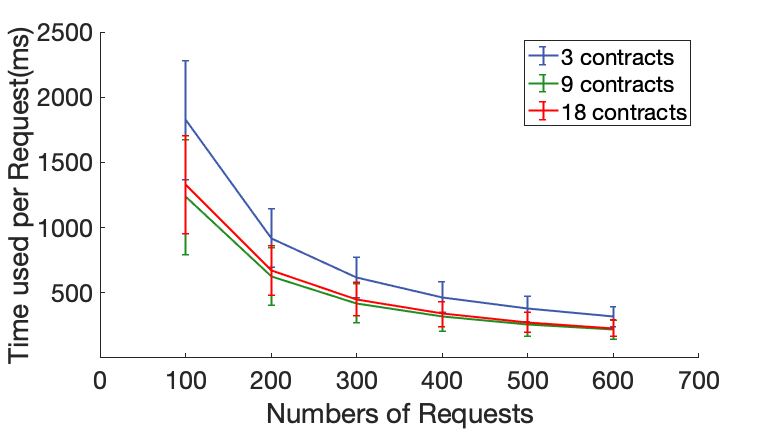}
        \caption{}
        \label{fig:time_cost_avg_std}
    \end{subfigure}
    ~ 
    \begin{subfigure}[t]{0.3\textwidth}
        \centering
        \includegraphics[height=1.2in]{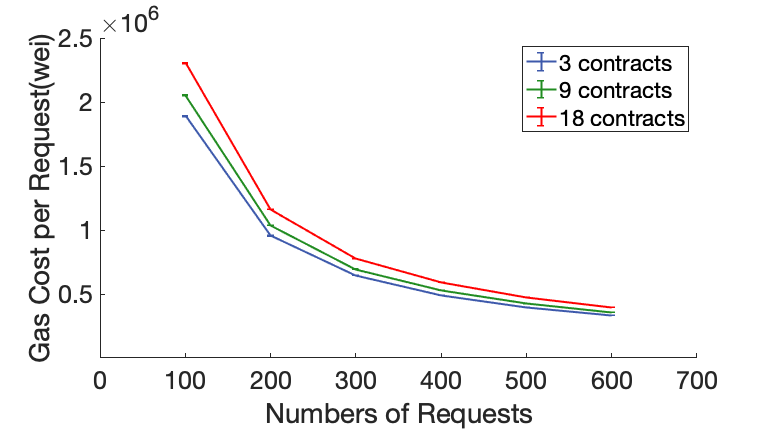}
        \caption{}
         \label{fig:gas_cost_avg_std}
    \end{subfigure}
    ~ 
    \begin{subfigure}[t]{0.25\textwidth}
        \centering
        \includegraphics[height=1.2in]{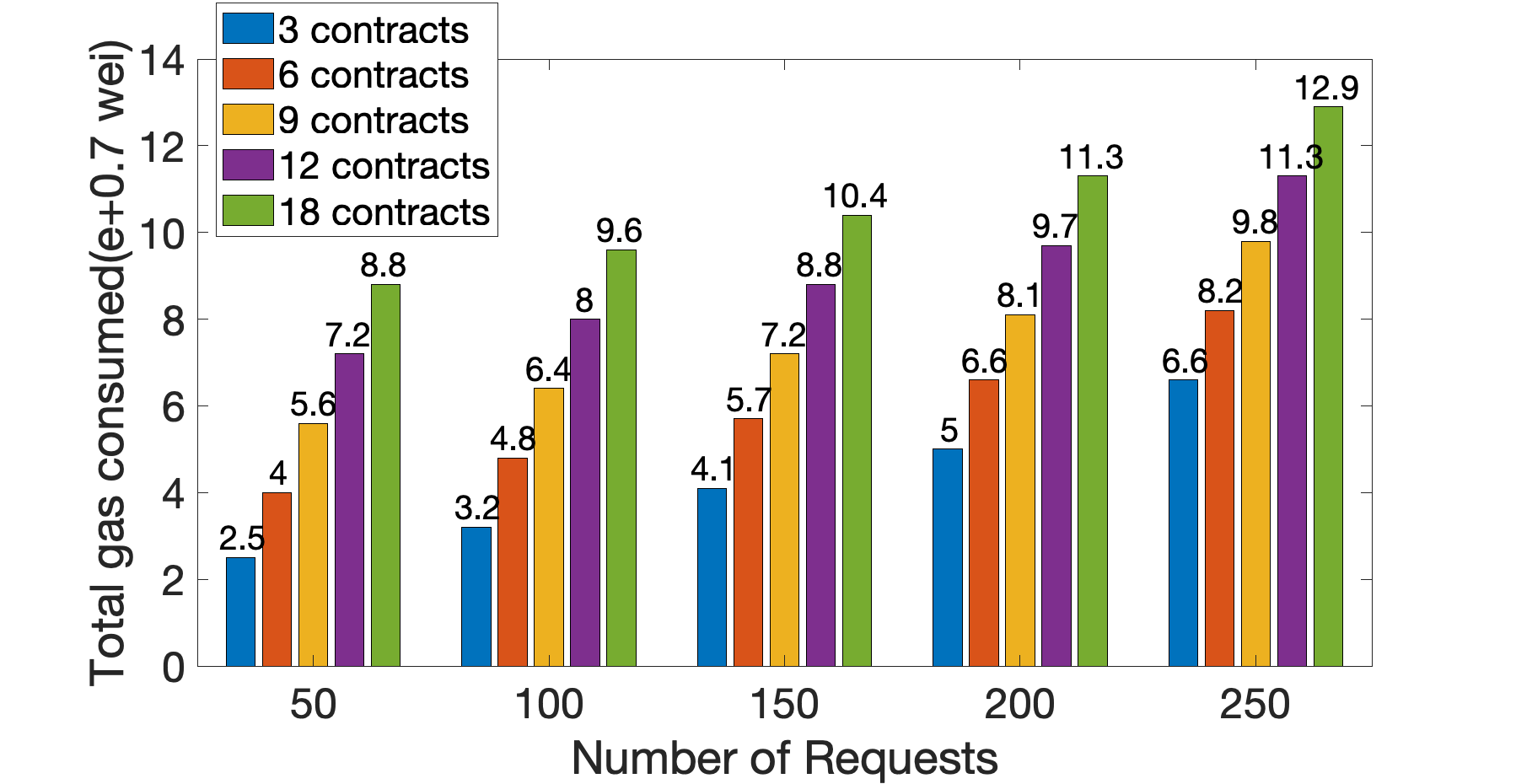}
        \caption{}
        \label{fig:total_gas_consumption}
    \end{subfigure}
\caption{(a) Average time cost for each request. (b) Average gas cost for each request. (c) Total gas consumed by system.}
\label{fig:fig}
    \vspace{-0.2cm}
\end{figure*}

\subsection{Security and Privacy}
Three leading technologies guarantee data security and personal privacy: decentralized database in the blockchain, automatic execution of smart contracts, and randomization of Bluetooth mac addresses. In addition, the localized data storage of user Bluetooth addresses and infection data enhances data security. 

(a) \textbf{Data Security: }
Our design guarantees that user data will not be manipulated. We choose blockchain as the system database, and its protocol\cite{xiao2020survey}\cite{vujivcic2018blockchain} stipulates that the current block must include the hash value generated from the previous block, so if an attacker manipulates a transaction in a block, he changes this block's hash value. Then, he must tamper with all the following blocks to secure his tampering, which will not be verified and accepted by other users.


Our system contains a smart contract group to handle all users' infection updates and their check-in requests for public locations. Smart contracts have the properties of being unmodifiable, requiring no supervision, and being automatically executed, so these distributed smart contracts ensure the same execution process and same output everywhere. So, smart contracts will not produce malicious results at certain computing nodes, which ensures the security of user data.

These designs also enhance the usability of our system. No individual user or computing note can be the centralized component to manipulate all data, which is the bottleneck in the system. Even partial smart contracts are generating malicious operations, the whole system is still accessible.

(b) \textbf{Identity Privacy: }

In our system, we use weak random Bluetooth mac addresses to replace the user's smartphone mac address\cite{8016185}\cite{gu2018bf} or IMEI number as their identities. This weak randomization protects the user's privacy because it is hard to associate these frequently changed random mac addresses with users' true identities in a short time, meanwhile, it costs less gas on Bluetooth communication. 

Combined with the Bluetooth protocol\cite{8016185}\cite{heydon2012bluetooth}\cite{martin2019handoff} in our design, we change the silent period\cite{1424677}, which refers to the gap time between discarding the old Bluetooth mac address and adopting the new mac address\cite{1424677}. And during this gap time, the Bluetooth device cannot decide to use the new or old address. Huang\cite{1424677} points out that changing the silent period's length obviously reduces the duration of the Bluetooth device being tracked, and thus the device will not be located easily. Thus, the modified randomization of Bluetooth address protects the user's privacy.

In our design, users' smartphones can scan and discover each other in a short distance, and then they exchange their Bluetooth mac addresses as temporary identities via Bluetooth advertisements directly without establishing a communication channel\cite{Bluetooth_Address}, which saves lots of battery consumption.

The design of weak randomization also solves another minor challenge. The larger quantity of random mac addresses, the better privacy protection. However, it leads to higher operating costs and network congestion. Therefore, we apply weak randomization of Bluetooth addresses to balance a sufficiently large quantity of Bluetooth random addresses for privacy protection and a relatively small quantity for lower system operating costs. 
\section{Evaluation}\label{Evaluation}

We build a prototype system on a private Ethereum platform and evaluate its performance on the average cost of processing requests and the total cost to operate the system. Our experiment simulates users’ daily contact and check-in activities by the Poisson distribution equation. 

\subsection{System Implementation}


Experiments run on a MacBook Pro with macOS version 10.14.5, Intel i5 CPU, and 8 GB memory. The smart contract group is developed by the language Solidity and deployed on a local private Ethereum simulated by Ganache software. Then, we use Python script to analyze the collected data.

Our experiments focus on primary variables affecting performance: 1). The number of $\mathtt{|U|}$ increasing from 100 to 600, with even intervals of 100, 2). the user's contact and check-in frequency $\mathtt{Freq}$ following Poisson distribution, and 3). smart contract group size $\mathtt{SCG.size}$. Moreover, we measure the average gas cost of all requests and the standard deviation on the average cost of ten rounds of experiments.

\subsection{Measure Avg Request Cost}
    
Based on three quantities of deployed smart contracts and six different numbers of requests increased from 100 to 600, we measure the average gas cost of all requests and the standard deviation on the average experiment cost of ten rounds.

Fig. \ref{fig:time_cost_avg_std} and \ref{fig:gas_cost_avg_std} shows that when the quantity of contracts increases from 3 to 18, and the number of requests increases from 100 to 600, both the average request time cost and gas cost are reduced by about 4 times. Time cost drops from 1,800 to 300 $\mathtt{ms}$ and gas cost drops from 2.3 to 0.5 million $\mathtt{wei}$. The time cost has a acceptable deviation and the variance of gas cost is tiny in Fig. \ref{fig:gas_cost_avg_std}. $\mathtt{Wei}$ is the smallest gas unit in Ethereum system and $\mathtt{ether}$ is $\mathtt{10^{18}}$ times the value of $\mathtt{wei}$. If 1 $\mathtt{ether}$ is worth $\mathtt{\$250}$, then the deviation of 500 $\mathtt{wei}$ is negligible.

\subsection{Evaluate System Overhead}

System overhead refers to the interaction costs between mobile services and users and operating costs for smart contracts measured by Ethereum gas.

Fig. \ref{fig:total_gas_consumption} shows that when the number of requests increases from 50 to 250, and the number of contracts increases from 3 to 18, the overall gas consumption increases linearly, considering the case of the same amount of requests with different amount of contracts, and the case of the same number of contracts with different amount of requests.

Based on the measurements, we conclude that this prototype system has good stability and scalability. With an increase in the number of requests and contracts, the request cost, which is the major overhead in the system operation, has a stable approach to a lower boundary, supporting the system's stability. Similarly, the system overhead is linear growth instead of an exponential one, which is acceptable.

\section{Related Work}\label{Related_Work}

In the field of contact tracing, MIT\cite{rivest2020pact}, Apple and Google\cite{Apple_Google_virus_tracking} have similar products and projects. However, their solutions are either a centralized system or have insufficient privacy protection for users. 

In terms of data security, smart contracts guarantee consistency in operational execution and obtain a consistent output, avoiding malicious data injection attacks. Jinyue\cite{song2020smart} presented a computing resource trading platform based on smart contracts, and we proposed a similar tree-structured smart contract group. However, their implementation focuses on matching users and completing resource trading, and our purpose of smart contracts is to process the infection status for locations. Moreover, our design of contracts hierarchy architecture contributes to updating the location infection status consistently.

Regarding privacy, some articles use differential privacy algorithms in IoT data processing\cite{lu2017lightweight}, but differential privacy methods will return a relatively accurate output with noise added. Their designs conflict with blockchain property because when storing data, the latter block must verify the correctness of the previous block's data and cannot tolerate deviations. 
\section{Conclusion}\label{Conclusion}

This paper designs this tracing and notification system based on blockchain and smart contract, which provides two services: location-based contact tracing and Bluetooth-based contact tracing, including notification function. Our system traces location infection status and records users' contact history locally. Also, it notifies users for infection contact. 

For users' privacy protection, we adopt random Bluetooth MAC address as a temporary identity for users. Due to battery limitations in smartphones, Bluetooth addresses are weakly randomized, and no communication channel is required for contact data exchanging. Besides, our hierarchy structured smart contracts ensure that each user can get the location's consistent infection results.. 

We simulate the interaction between users and our prototype system, and evaluate its performance, including gas consumption, operating stability, and request processing speed. In the simulated environment, our system has good scalability and good stability. We expect to have real data about user contact records to evaluate our system in the future.

\section{Future Work}\label{Limitation and Future_Work}

Our architecture may enrich the services with health analysis functions, such as estimating the probability of being infected based on the user's visiting and contact history.  Currently, our design can only remind users that they have been exposed to a virus environment and have contacted patients. It lacks more prosperous analysis functions, such as estimating the probability of being infected and statistical analysis of an infection frequency within a location. 

In the future, we would explore the relationship between the Received Signal Strength Indication $\mathtt{RSSI}$ and the possibility of being infected by a virus. The research and experiments made by MIT\cite{MIT_airborne}\cite{MIT_Sneeze} show that the closer a user is to the infected person, the easier he is to be exposed to more viruses and be infected. Also, Jung\cite{Jung2013DistanceEO} found that the distance between two users can be measured by the $\mathtt{RSSI}$ of Bluetooth. If these factors can be correlated, this will enhance the notification and tracing mechanism. We may extend this system to a more generic infectious disease protection research.

\bibliographystyle{ieeetr}
\bibliography{main}

\end{document}